\documentclass[12pt]{article}
\usepackage{amssymb}
\usepackage{epsfig,color}
\usepackage{pstricks,graphicx,epsfig,color,amssymb,amsmath,amscd}
\usepackage{cite}
\usepackage[]{graphicx}
\usepackage{makeidx}
\usepackage{amsmath}
\usepackage{nccmath}
\usepackage{setspace}
\usepackage{ulem}
\usepackage{tensor}
\usepackage{bm}

\usepackage[]{caption}
\renewcommand\rho{\varrho}

\newcommand{\be}{\begin{eqnarray}}
\newcommand{\ee}{\end{eqnarray}}
\newcommand{\rar}{\rightarrow}

\begin{document}

\title{The highest energy cosmic rays from superheavy dark matter particles }
\author{E.\,V. Arbuzova}
\date{}

\maketitle
\begin{center}

{\it arbuzova@uni-dubna.ru}\\

{Department of Higher Mathematics, Dubna State University,\\ 
Universitetskaya st.\,19, Dubna, 141983 Russia}\\
{Department of Physics, Novosibirsk State University,\\ 
Pirogova st.\,2, Novosibirsk, 630090 Russia}\\
\end{center}

\begin{abstract} 
It is commonly accepted that high energy cosmic rays up to $10^{19}$ eV
can be produced in catastrophic astrophysical processes. However the
source of a few observed events with higher energies remains
mysterious. We propose that they may originate from decay or
annihilation of ultra heavy particles of dark matter.  Such particles
naturally appear in some models of modified gravity related  to
Starobinsky inflation. 
\end{abstract}

\maketitle

\section{Introduction}

There is no single opinion on the origin of Ultra High Energy Cosmic Rays (UHECR). The traditional approach is that
the high energy cosmic rays are created by astrophysical sources such as active galactic nuclei, Seyfert galaxies, or potentially hypernovae.
For a recent review, see Ref. \cite{Kachelriess:2022phl}.
Other intriguing possibilities include cosmic ray emission from topological defects \cite{Sigl:1996ak,Semikoz:2007wj}, the decay of superheavy particles in supersymmetric theories influenced by instanton and wormhole effects \cite{Berezinsky:1997hy,Kuzmin:1997jua,Birkel:1998nx}, binary neutron star mergers, and similar phenomena.

There are two separate energy ranges for ultra high energy cosmic rays. Cosmic rays with energies $E \lesssim 10^{20}$ eV can potentially originate from stellar processes, where stellar material is accelerated during catastrophic events. These cosmic rays typically contain a significant fraction of nuclei. On the other hand, cosmic nuclei with energies $E \gtrsim 10^{20}$ eV are difficult to produce by stellar mechanisms. Such extremely energetic cosmic rays could, in principle, be generated through the decay of heavy particles.

\begin{figure}[htbp]
		\vspace{-0.5cm}
		\begin{center}
		\includegraphics[scale=0.35,angle=-90]{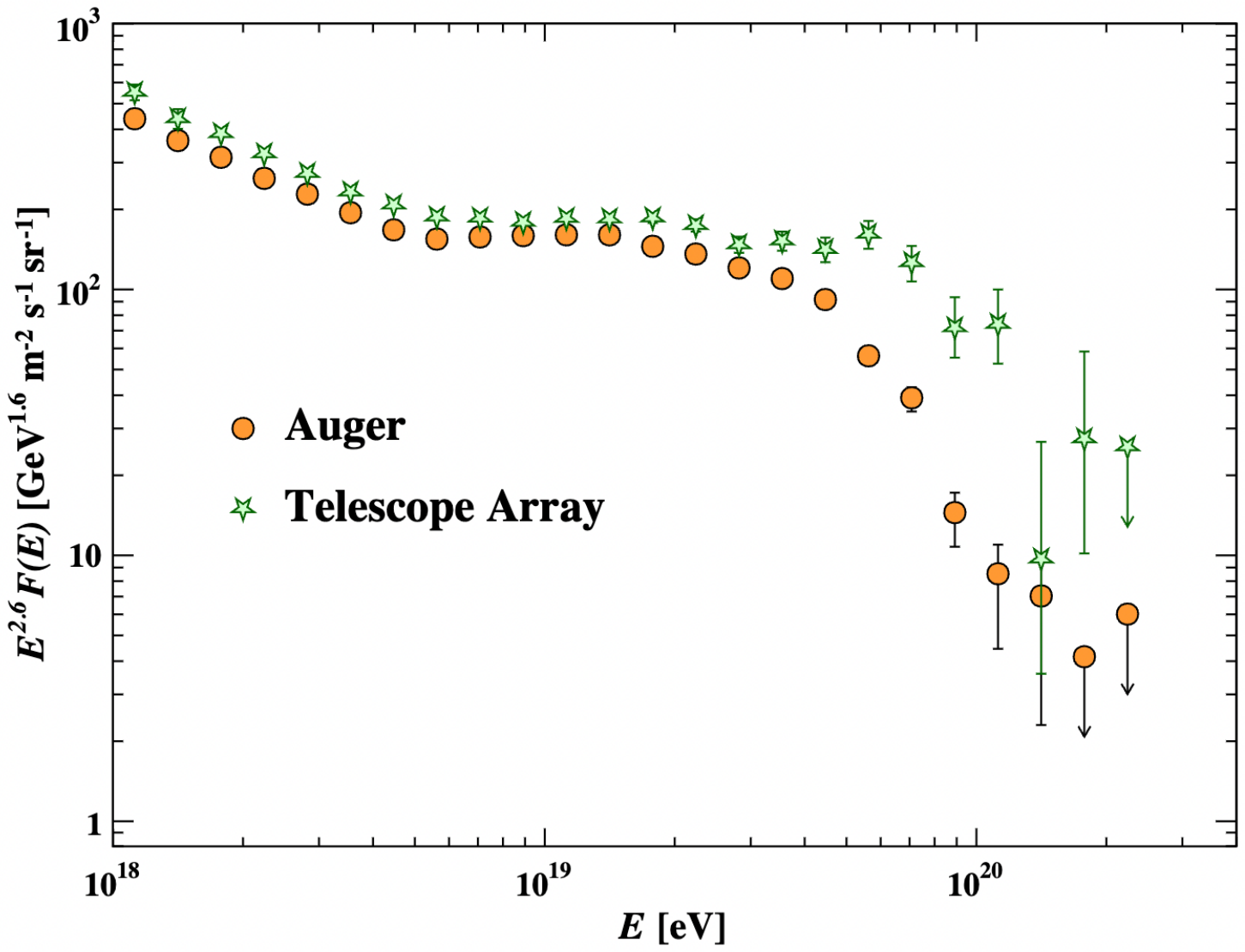}
		\vspace{-3mm}
       \end{center}
	\caption{UHECR flux observations, from Fig. 30.10 of Ref.~\cite{ParticleDataGroup:2020ssz}. 
	}
	\label{fig_UHECR}
\end{figure}

Fig.~\ref{fig_UHECR} shows the UHECR spectrum, measured by the Pier Auger and Telescope Array observatories~\cite{ParticleDataGroup:2020ssz}.  We are particularly 
interested in the very end of the spectrum, specifically  the part corresponding to the highest energies, in the region of $10^{20}$ eV. 
It is clearly visible that there are several events with energies exceeding $10^{20}$ eV. Cosmic rays of such high energies are sometimes called 
as Extremely High Energy Cosmic Rays (EHECR). The origin of cosmic rays with such extremely high energies remains a mystery and the efforts 
of many scientists are focused on finding a solution to this problem. 

We suggest other possible sources of EHECR at energies $E \gtrsim 10^{20} $ eV: 1) the annihilation of superheavy Dark Matter (DM) particles created by the scalaron 
decay in $R^2$-modified gravity; 2)~the decay of such superheavy particles via virtual black holes in multidimensional gravity. 

We assume that superheavy dark matter particles are produced by the oscillating curvature scalar in the model of the Starobinsky inflation~\cite{Starobinsky:1980te} 
with the action: 
\be
S (R^2) = -\frac{M_{Pl}^2}{16\pi} \int d^4 x \sqrt{-g}\,\left[R- \frac{R^2}{6M_R^2}\right], 
\label{action-R2}
\ee
where $M_{Pl} = 1.22 \cdot 10^{19}$ GeV is the Planck mass\footnote{ Throughout the paper, we use
 the natural system of units, where the speed of light $c=1$, the Boltzmann constant $k=1$, and the reduced 
Planck constant $\hbar = 1$.
The gravitational coupling constant is given by $G_N = 1/M_{Pl}^2$, where the Planck mass is 
$M_{Pl} = 1.22 \cdot 10^{ 19}\,\rm{GeV} = 2.17\cdot 10^{-5}\,{\rm g} $.
For reference, the following conversions apply:
1~GeV$^{-1} = 0.2\cdot 10^{-13}$ cm, 1sec $ = 3\cdot 10^{10}$ cm, and 1yr = $3.16\cdot 10^{7} $ sec.}.    
The non-linear term in the action leads to the appearance of a new dynamical  scalar degree of freedom, 
$R(t)$, known as {\it scalaron}. The parameter $M_R$
represents the scalaron mass. Angular fluctuations of the cosmic microwave background radiation (CMBR) imply the following value for the scalaron mass:
$M_R \approx 3 \cdot 10^{13} $ GeV~\cite{Faulkner:2006ub}. 

The calculations of the scalaron decay probabilities were performed in our paper~\cite{Arbuzova:2021oqa},
as well as in several others. In all these works, it was assumed that the masses of the decay products were
much smaller than the scalaron mass, $M_R$.
 In our papers \cite{Arbuzova:2018apk,Arbuzova:2020etv,Arbuzova:2021etq},  
 the  production of superheavy carriers of dark matter 
via scalaron decays into particles 
with masses up to $M \lesssim 10^{12}$ GeV was studied. In all cases  the condition $(2 M/M_R)^2 \ll 1$ was satified. 

We now investigate an alternative source of high-energy cosmic rays, namely, the annihilation and decay of superheavy dark matter particles. We assume that these heavy dark matter particles are fermions directly produced by scalaron decays. Previous calculations of scalaron decay probabilities were performed in the limit of low masses for the decay products. For example, the width of scalaron decay into a fermion-antifermion pair was found to be:
 \be
  \Gamma_{m_f} = \frac{m_f^2 M_R}{6 M_{Pl}^2},
  \label{Gamma-f-old}
  \ee
 where $m_f$ is the fermion mass.
   For the decay width given by Eq.~\eqref{Gamma-f-old},  the energy density of heavy fermions would be much larger than the averaged 
   cosmological density of dark matter:
 \be
 \rho_{DM} \approx 1 \ \text{keV/cm}^3.
 \label{rho-DM}
 \ee
 
 The probability of the decay would be strongly suppressed 
 if the fermion mass $M_f$ is extremely close to $M_R/2$:
  \be 
   \Gamma_f = \frac{M_f^2 M_R}{6 M_{Pl}^2}
    {\sqrt{1- \frac{4M_f^2}{M_R^2}}}, \ \ \ M_f \sim \frac{M_R}{2}.
 \label{Gamma-f}
 \ee
 The phase space factor  $(1- 4M_f^2/M_R^2)^{1/2}$  allows for the energy density of the presumed dark matter particles
  $f$  to be arranged such that it matches the observed dark matter density: $\rho_f \approx \rho_{DM}$.

The paper is organized as follows. In Sec.~\ref{s-annih}, we consider the flux of cosmic rays resulting from the annihilation of superheavy dark matter particles. 
Several regions of the universe where such annihilation could take place are studied. These include: 
\begin{itemize}
\item
The entire universe under the assumption 
of homogeneous dark matter energy density.
\item
The high density dark matter clump in the Galactic Center.
\item
The cloud of dark matter in the Galaxy with realistic energy density distribution of dark matter.
 \end{itemize} 
In Sec.~\ref{s-decay}, we show that stable by assumption dark matter particles could decay via interactions with virtual black holes. 
In multidimensional gravity, where the scale of gravitational interaction is smaller, the decay of these ultra-massive particles 
can provide a noticeable contribution to the flux of UHECR. In the last section we conclude.

\section{Annihilation of superheavy dark matter particles \label{s-annih} }

Let us consider the flux of cosmic rays resulting from the annihilation of heavy dark matter particles, which were created during the scalaron decay process (for details, see Ref.~\cite{Arbuzova:2024uwi}).

The flux of high energy particles is determined by the cross-section for the annihilation of heavy fermions:
\be 
\sigma_{ann} v \sim \alpha^2 \,g_*/M_f^2,
\label{cr-sec} 
\ee
where $v$ is the center-of-mass velocity,  $\alpha$ is the coupling constant, $\alpha \sim 10^{-2}$, and $g_*$ is the number of the open annihilation 
  channels,  $g_* \sim 100$.  With fermion mass $M_f = 1.5 \cdot 10^{13}$ GeV we estimate  $\sigma_{ann} v \sim 2 \cdot 10^{-56} \text{cm}^2$. 
  Below, following our work  in Ref.~\cite{Arbuzova:2024uwi}, 
  we propose a way to enhance the efficiency of the annihilation. 
  
  The rate of the decrease of the $f$-particle density per unit of time and volume is equal to:
\be
\dot n_f = \sigma_{ann} v n_f^2 = \alpha^2 g_* n_f^2/M_f^2, 
\label{dot-n-f}
\ee

We assume that the annihilation is sufficiently slow, such that the number density 
$n_f$ significantly change over the age of the universe.

The annihilation of heavy $f$-particles results in a continuous contribution to the rate of cosmic ray production per unit time and unit volume, given by:
\be 
\dot \rho_f = 2M_f \dot n_f.
\label{dot-rho-f3}
\ee
The results in Eqs. \eqref{dot-n-f} and \eqref{dot-rho-f3} are valid for the total flux integrated over particle energy.

To compare our results with observational data, we need to determine the energy distribution of cosmic ray particles produced in the process of
$f \bar f$ - annihilation. We postulate that the differential energy 
spectrum of the number density flux, $\dot n_{PP} (E)$, of the produced particles  is stationary 
and adopts a form that we consider reasonable:
\be 
\frac{d \dot n_{PP} (E)}{dE}  =  {\mu^3}\, \exp \left[ -\frac{(E - { 2} M_f/\bar n)^2}{\delta^2}\right] \theta(2M_f-E) .
\label{dn-dE}  
\ee
Here, $\mu$ is a normalisation factor, with dimension of mass (or equivalently inverse length),
which will be determined later, and $\bar n$ is the average number of particles created in the  
process of $f\bar f$ - annihilation. This distribution ensures that the maximum energy of the annihilation products is
$E_{max} = 2M_f$ and the average energy per particle is approximately
$\bar E \approx 2M_f/\bar n$, if the width of the distribution, $\delta$, is  sufficiently small. 

The contribution from the annihilation of heavy particles to the cosmic ray flux is given by:
\be 
\frac{d \dot \rho_{PP} (E)}{dE} = E\, \frac{d \dot n_{PP} (E)}{dE}.
\label{drho-de} 
\ee
Correspondingly, the total energy density flux of the produced particles, with the number density spectrum given by Eq.~\eqref{dn-dE}, is:
\be
\dot \rho_{PP} =  \int_0^{2M_f}  E  \left(\frac{d\dot n_{PP}(E)}{dE}\right)\,dE\,  \approx \sqrt{\pi}\, { {\mu^3}}\,\bar M  \delta \,, \ \ \bar M = { 2} M_f /\bar n.
\label{dot-rho-f-2}
\ee
We assume, that $\dot \rho_{PP} = const$, since the observed flux of the cosmic rays is stationary. 

Taking $n_f = \rho_{DM} / {2 M_f}$ and $M_f = 1.5\cdot 10^{13} $ GeV we calculate the total energy rate of cosmic rays produced by
 $f\bar f$ -annihilation as follows:
\be 
\dot\rho_f^{(ann)} = 2M_f \dot n_f = 2 \alpha^2 g_* n_f^2/M_f= { 1.48}\cdot 10^{-54} \,\text{GeV}^{-1} \text{cm}^{-6}.
\label{dot-rho-f-ann}
\ee
We determined the normalization factor ${\mu^3}$ based on the condition of equality between  $\dot\rho_{PP}$ from Eq. (\ref{dot-rho-f-2}) and 
$\dot\rho_f^{(ann)}$ from Eq.~(\ref{dot-rho-f-ann}):  
\be
{\mu^3} = \frac{1.48\cdot 10^{-54} \bar n}{2\sqrt{\pi} {\rm GeV \cdot cm^6} M_f \delta} =
{\frac{2.2 \cdot 10^{-109}\, \bar n}{{\rm cm^3}} \,\left(\frac{\rm GeV}{\delta}\right).}
\label{C}
\ee

Let us estimate the energy flux of the products of  the annihilation of dark matter particles 
"in the entire Universe" and reaching Earth's detectors, assuming that dark matter in the Universe is distributed uniformly and isotropically. 

We calculate the flux of cosmic rays for the spherical volume of radius $R$ assuming a homogeneous distribution of $f$-particles.
We take  $R_{max} \approx10^{28}$cm, as beyond this distance the redshift cutoff becomes significant. 
Finally, we estimate the energy flux from the entire Universe, assuming (unrealistically) a homogeneous distribution of dark matter, as follows.
The flux created by a source $S$ from the spherical layer with radius  $R$  and thickness $\Delta R$ is given by: 
\be 
\Delta L = \frac{S}{4 \pi R^2}\times 4\pi R^2 \Delta R = S \Delta R.
\label{Delta-F}
\ee
Integrating over the homogeneity scale, we obtain the total flux:
\be
L _{hom}= S R_{max}.
\label{F}
\ee
In the case under consideration, the source term, $S_{hom}$, is expressed as:
\be
S_{hom} = \frac{d \dot n_{PP}}{dE}, 
\label{S-of-rho}
\ee
where $(d \dot n_{PP}/dE)$ is given by Eq. (\ref{dn-dE}) with $\mu^3$ determined by expression (\ref{C}).
Note that the dimension of $S$ is $[eV^3] $ or $[cm^{-3}]$, and hence the dimension of $L$ is $[cm^{-2}]$. 

Now, we can calculate the contribution to the
flux of high energy cosmic rays, arising from  $f \bar f$ annihilation, as:
\be 
L _{hom}=  \frac{2.23 \cdot 10^{-109}\cdot 10^{28}\, \bar n}{cm^2}\,\left(\frac{\rm{GeV}}{\delta} \right)
\exp \left[ -\frac{(E - { 2} M_f/\bar n)^2}{\delta^2}\right] \theta(2M_f-E).
\label{L}
\ee
A crude order-of-magnitude estimate of $L$ from Eq. \eqref{L}, assuming $\bar n = 10^3$ and $\delta \sim 1$ GeV, gives
$L_{ hom}\sim 10^{-78}$\,cm$^{-2}$. The observed flux, which can be extracted from Fig.~\ref{fig_UHECR} (for details, see Ref. \cite{Arbuzova:2024uwi}),   
is approximately a few times  $10^{-55}$\,cm$^{-2}$. Thus, the observed flux exceeds
 the theoretical prediction by 23 orders of magnitude. The smallness of this result is explained by extremely week annihilation cross-section 
 given in Eq.~(\ref{cr-sec}).

However, the annihilation can be significantly enhanced due to the resonance process of  $f\bar f$- transition to the scalaron, as $2 M_f $ is very
close to $M_R$. Resonance effects in dark matter particle annihilation have been discussed in Refs.~\cite{Griest:1990kh,Gondolo:1990dk}. Eq.~(\ref{dot-n-f})  is valid
for S-wave annihilation with an energy-independent cross-section. For an arbitrary dependence of the cross-section on the center-of-mass energy
squared, $s = (p_f+p_{\bar f})^2$, the average value of $\sigma_{ann} v$ is calculated in Ref.~\cite{Gondolo:1990dk}:
\be
\langle \sigma_{ann} v \rangle = \frac{1}{8 M^4_f T [K_2(M_f/T)]^2} \int_{4M_f^2}^\infty ds\,(s- 4 M^2_f) \sigma_{ann} (s) \sqrt{s} 
K_1\left(\frac{\sqrt{s}}{T}\right) ,
\label{sigma-aver}
\ee
where $T$ is the cosmic plasma temperature, and $K_{(1,2)}$ are the modified Bessel functions. 
Since $x =M_f/T \gg 1$ we can use the asymptotical limit  of  $K_n (x) \approx \sqrt{\pi/(2x)}\, e^{-x}$. 
Substituting this approximation, we obtain:
\be
\langle \sigma_{ann} v \rangle = \frac{4}{\sqrt{\pi}} \sqrt{\frac{T}{M_f}}\int_0^\infty dz z e^{-z} \sigma_{ann}(z)  
\label{sigma-v-2},
\ee
where the dimensionless variable $z$ is defined as $s = 4M^2_f (1+T z /M_f)$. 

In our case, the cross-section exhibits a resonance due to the intermediate scalaron state in f anti-f - annihilation, 
 as the scalaron mass is very close to the sum of the masses of 
 $f$ and $\bar  f$. According to Ref.~\cite{Griest:1990kh}, the resonance cross-section is given by:
\be
\sigma_{ann}^{(res)} v = \frac{\alpha ^2 s }{(M_R^2 - s)^2 + M_R^2 \Gamma_R^2},
\label{sigma-res}
\ee
where $M_R = 3\cdot 10^{13} $ GeV is the scalaron mass, and $\Gamma_R $ is its decay width, equal to  (see Eq. \eqref{Gamma-f-old})
$\Gamma_R = M_f^2 M_R/(6M_{Pl}^2)$~\cite{Arbuzova:2021oqa,Arbuzova:2018apk}.

Now, for the thermally averaged resonance cross-section derived from Eqs.~\eqref{sigma-v-2} and \eqref{sigma-res}, we obtain:
\be
\langle \sigma_{res} v \rangle =  \int_0^\infty dz z e^{-z} \frac{ \alpha^2 s }{(M_R^2-s)^2 + M_R^2 \Gamma_R^2 } 
 =
 \frac{\alpha^2}{M^2_{ R}}\int^\infty_0 \frac{ dz z e^{-z}}{\gamma^2 +\eta^2 z^2} ,
\label{sigma-res-1}
\ee
where $\gamma^2 =\Gamma_R^2/M_R^2 = 1/36\,(M_f/M_{Pl})^4 \approx 6.7\cdot 10^{-26}$, and 
$\eta^2 = (T/M_f)^2 \approx 2.45\cdot 10^{-52} $. Here, we used $T=T_{CMB} = 2.7K = 2.35 \cdot 10^{-4}$ eV and $M_f = 1.5 \cdot 10^{13}$ GeV. 

Thus, the term $\eta^2 z^2$ can be neglected, leading to the conclusion that the resonance cross-section is 26 orders of magnitude higher than the previous estimate. Consequently, the contribution to the flux of cosmic rays may reach a sufficient level to explain the origin of ultra high energy cosmic rays with  $E \gtrsim 10^{20}$ eV.

The effect is even stronger in the case of 
$f\bar f$-annihilation in denser regions of the Galaxy with the realistic distribution of dark matter. 

Let us estimate the flux of cosmic rays originating from dark matter annihilation in the Galactic Center, where the local DM density is significantly higher than the average cosmological density \cite{Sofue:2020rnl}:
\be
\rho_{GC} = 840 \ \text{GeV/cm}^3.
\label{rho-GC}
\ee
This value exceeds the average DM density by 9 orders of magnitude. Since the flux of cosmic rays from DM annihilation is proportional to the square of the DM particle density, smaller regions with higher local DM density can produce a significantly larger flux of cosmic rays compared to the average density regions.

In Eq.~\eqref{F}, the flux of cosmic rays $L$
from DM annihilation in the entire galaxy is presented under the (unrealistic) assumption of a homogeneous distribution of dark matter. 
The obtained result should be rescaled as follows.
\begin{enumerate}
\item
Multiply by the square of the ratio of the DM density in the Galactic Center to the average cosmological DM density
since the annihilation rate is proportional to $n_f^2$, see Eq.~(\ref{dot-n-f}).
\item
Multiply by the volume of the high-density clump in the Galactic Center, $4\pi r_{cl}^3/3$, where $r_{cl}$ is the radius of the clump.
\item
Divide by the area of a sphere at the distance $d_{gal}$ from the Galactic Center, $4\pi d_{gal}^2$.
\end{enumerate}
Thus the following rescaling is to be done:
\be 
{ L_{GC}} = L_{hom} \times \left(\frac{n_{GC}}{\bar n_{DM}}\right)^2 \frac{r_{cl}^3/(3\,d_{gal}^2)}{R_{max}} { \approx 10^3 \,  L_{hom}},
\label{L-rescale}
\ee
where $L_{hom}$ is determined by Eqs.~(\ref{F}), \eqref{S-of-rho}, (\ref{L}).  The factor
$d \dot n_{pp}/{dE}$,  entering these expressions, is given by Eq.~(\ref{dn-dE}) and $R_{max} =10^{28}$ cm. 

We assume that the size of this
high density clump in the Galactic Center is approximately $r_{cl} =10 \rm{pc} \approx 3\cdot 10^{19}$ cm and its distance
to Earth is  $d_{gal} =$ 8 kpc = 2.4 $\cdot 10^{22}$ cm.  Thus, the flux could be increased by the factor $1.1\times 10^3$.

To conclude this section, let us consider the flux of cosmic rays resulting from the annihilation of dark matter with a realistic distribution in the Galaxy.

We adopt the commonly accepted shape of the dark matter distribution~\cite{Gunn:1972sv}:
 \be
 {\displaystyle \rho (r)=\rho _{0}\left[1+\left({\frac {r}{r_{c}}}\right)^{2}\right]^{-1}} \equiv \rho_0 q(r) ,
 \label{dm-of-r}
 \ee
 where $\rho _{0}$ denotes the finite central density and $r_{c}$ is the core radius. For the sake of estimation, we assume $r_{c}=1$~kpc and
 calculate $\rho _{0}$ under the condition that at the position of the Earth
 at $r = l_\oplus= 8$ kpc  the density of dark matter is approximately $ \rho (l_\oplus)\approx  0.4\, {\rm GeV/cm}^3$~\cite{Salucci:2010qr}. Hence, we find: 
 \be  
 \rho_0 = 65 \rho ( l_\oplus) =
 26 \,{\rm GeV/cm}^3.
  \label{rho-center}
 \ee
This value exceeds the average cosmological dark matter density,
$\rho_{DM} = 1$~keV/cm$^3$  by a factor of  $2.6 \times 10^7$.

Let us consider the annihilation of DM particles at a point specified by the radius vector $\vec r$, 
expressed in spherical coordinates $(r, \theta, \phi)$, directed from the Galactic Center. The distance of this point to  Earth is given by:
\be
d_\oplus =\sqrt{ (\vec{l_\oplus} + \vec{r})^2 } =\sqrt{ r^2 + l_\oplus^2 - 2 r\,l_\oplus \cos{\theta}}.
\label{d-plus}
\ee

As  done above, we recalculate the flux of cosmic rays by rescaling Eq.~\eqref{L-rescale} 
using the ratio of the dark matter density, given by
(\ref{dm-of-r}) and (\ref{rho-center}), to the DM density in the Galactic Center. Additionally, we include the integral over the DM distribution up to $R_{max}$.
Thus, for a realistic DM distribution in the Galaxy, we obtain:
\be 
L_{real} = L_{GC} \left( \frac{26 {\rm GeV}}{840 {\rm GeV}}\right)^2 \frac{{3} d_{gal}^2}{ r_{cl}^3 } J,
\label{L-real}
\ee
where J is the integral over DM distribution:
\be
J= \int \frac{d^3 r q(r)}{d_\oplus ^2} 
= 2\pi  \int \frac{dr r^2 q(r)d\cos \theta}{r^2 + l_\oplus^2 - 2 r  l_\oplus\cos \theta} = 
2\pi \int \frac{dr r q(r)}{ l_\oplus} \ln \frac{l_\oplus+r}{l_\oplus-r}.
\label{J}
\ee
After a change of variables, $r = x l_\oplus$, the integral is reduced to the following expression and is evaluated numerically:
\be
J = 2\pi\, l_\oplus \int_0^1 dx x \left( 1 + 64 x^2\right)^{-1} \ln \frac{1+x}{1-x} = 0.2 \,l_\oplus . 
\label{J-2}
\ee
Thus, we obtain $L_{real} = 3\cdot 10^5 L_{GC} $. This value is significantly larger than the flux originating from the dense Galactic Center 
and requires much weaker amplification through resonance annihilation (Eq.~\eqref{sigma-res}).

\section{Decay of superheavy dark matter particles through virtual black hole in multidemensional gravity  \label{s-decay}} 

In this section we consider another possibility for creation of ultra high energy cosmic rays: through the decays of superheavy dark matter particles 
produced by oscillating curvature. This problem is discussed in detail in the works \cite{Arbuzova:2023dif,Arbuzova:2024ahj}, here we briefly 
focus on main points.  

Dark matter particles are usually assumed to be absolutely stable. However, in 1976, Ya.~B.~Zeldovich proposed a mechanism
whereby the decay of any presumably stable particles could occur through the creation of virtual black holes~\cite{Zeldovich:1976vq,Zeldovich:1977be}.
The rate of proton decay, as calculated in the framework of canonical gravity with an energy scale equal to $M_{Pl}$, is extremely small. The corresponding lifetime of the proton far exceeds the age of the universe ($t_U \approx 1.5\cdot 10^{10}$  years). However, the smaller scale of gravity and the huge mass of dark matter particles can both lead to a strong amplification of the Zeldovich effect.

We are particularly interested in superheavy dark matter particles with masses around $M_ X \sim 10^{12}$ GeV. 
These particles may decay through the formation of virtual black holes with lifetimes several orders of magnitude longer than the universe age. 
 The decay of such particles could make a significant contribution to the generation of ultra high energy cosmic rays.

We consider the model proposed in Refs.~\cite{Arkani-Hamed:1998jmv,Antoniadis:1998ig}, where the observable universe, 
containing the Standard Model particles, is confined to a 4-dimensional brane embedded in (4+$d$)-dimensional bulk. 
In this framework, gravity propagates throughout the bulk, while all other interactions remain restricted to the brane.
In such scenarios, the Planck mass, $M_{Pl}$, becomes an effective long-distance 4-dimensional parameter, and its relationship with the fundamental gravity scale,
$M_{\ast}$, is given by: 
\begin{eqnarray}
M_{Pl}^2\sim M_{\ast}^{2+d}R_*^d ,
\label{M-ast}
\end{eqnarray}
where $R_*$ is the size of the extra dimensions. The size $R_*$ can be expressed as: 
\be
R_{*} \sim \frac{1}{M_*}\left(\frac{M_{Pl}}{M_*}\right)^{2/d}.
\label{R-of-M}
\ee
For the purposes of our application we choose
$M_* \approx 3\cdot10^{17}$ GeV, leading to $R_*\sim 10^{(4/d)}/M_* > 1/M_*$.

The width of proton decay via a virtual black hole into a positively charged lepton and a quark-antiquark pair, 
in the framework of the multidimensional gravity model, was calculated in Ref.~\cite{Bambi:2006mi} and is given by the following expression:
\be 
\Gamma(p\rar l^+ \bar q q) = 
\frac{m_p\,\alpha^2}{ 2^{12} \, \pi^{13}}
\left(\ln \frac{M_{Pl}^2}{m_q^2}\right)^2 \,
\left(\frac{\Lambda}{M_{Pl}}\right)^6 \,
\left(\frac{m_p}{M_{Pl}}\right)^{4+\frac{10}{d+1}}\, 
\int_0^{1/2} dx x^2 (1-2x)^{1+\frac{5}{d+1}}.
\label{gamma-p}
\ee
 Here, $m_p \approx 1$ GeV is the proton mass, $m_q \sim 300$ MeV is the constituent quark mass,  
 $\Lambda \sim 300$ MeV is the QCD scale parameter, $\alpha = 1/137$ is the fine structure constant,
 and $d$ denotes the number of "small' extra dimensions. The QCD coupling constant $\alpha_s$ is assumed to be equal to unity.
 
We applied the above result to the process  $X\rar L^+ \bar q_* q_*$, assuming that the heavy dark matter $X$-particle, with mass $M_X \sim10^{12}$ GeV,  
consists of three heavy quarks, $q_*$, with comparable masses.  
The parameter $\Lambda_* $ is left as a free variable. Substituting $M_*$ in place of $M_{Pl}$,
the life-time of $X$-particles can be evaluated using Eq.  \eqref{gamma-p}.  
In this evaluation, we make the following substitutions:
\begin{itemize}
\item
$\alpha_*= 1/50$ instead of $\alpha = 1/137$, 
\item
$M_X=10^{12}$ GeV  
instead of $m_p$, 
\item
the mass of the constituent quark $m_{q_*} = 10^{12}$~GeV,
\item
  $d=7$ for the number of extra dimensions. 
 \end{itemize} 
  Thus, the life-time  of the $X$-particle can be expressed as: 
\be \label{tau-X}
\tau_X = \frac{1}{\Gamma_X} 
\approx {6.6\times 10^{-25} \rm{s} \, \cdot\frac{2^{10} \pi^{13}}{\alpha_*^2}} \left(\frac{\rm{GeV}}{M_X}\right)
\left(\frac{M_*}{\Lambda_*}\right)^6  \left(\frac{M_*}{M_X}\right)^{4+\frac{10}{d+1}}
\left( \ln \frac{M_*}{m_{q_*}} \right)^{-2} I(d)^{-1},
\ee
where we took {1/GeV = $6.6\times 10^{-25} $\rm{s}} and
\be
I (d) =\int_0^{1/2} dx x^2 (1-2x)^{1+\frac{5}{d+1}}, \,\,\, I(7) \approx 0.0057.
\label{i-of-d}
\ee
Now, all parameters, except for $\Lambda_*$, are fixed: $M_*=3\times 10^{17}$~GeV,
$M_X = 10^{12}$ GeV, $m_{q_*} \sim M_X$. The life-time of X-particles can be estimated as: 
\be 
\tau_X \approx 7\times 10^{12}\,\,  {\rm years} \left(10^{15}\,\rm{GeV} /\Lambda_* \right)^6 \ \ \ vs \ \ \ 
t_U \approx 1.5\times 10^{10}\ {\rm years}. 
\label{tau-x-2}
\ee
A slight variation of $\Lambda_*$ near {$10^{15}$ GeV} allows to fix the life-time of the dark matter 
X-particles in the interesting range. These particles would be stable enough to behave as cosmological dark matter, 
while their decay could contribute significantly to the production of cosmic rays at ultra high energies. 

\section{Conclusions} 

\begin{itemize}
\item
In $R^2$-modified gravity the viable canditates for dark matter particles could be very heavy, up to $M \sim 10^{13}$ GeV. 
\item
The annihilation and decay of such superheavy particles are promising sources of extremely high energy cosmic rays, which 
are difficult to explain using canonical astrophysical mechanisms.
\item
The contribution to the flux of cosmic rays originated from different cosmological environment (e.g. DM clumps in the Galactic Center)
might be sufficient to explain the origin of UHECR with $E \gtrsim 10^{20}$ eV. 
\item
The flux of UHECR could be significantly enhanced in the case of resonance annihilation of superheavy 
DM particles with masses close to a half of the scalaron mass.
\item
DM particles are assumed to be stable with respect to conventional particle interactions. However, they could decay through the formation of virtual black holes. In the framework of high-dimensional gravity, the lifetime of such quasi-stable particles may exceed the age of the universe by several orders of magnitude.
\item
The mechanisms discussed may provide an explanation for the origin of UHECR observed by the Pierre Auger Observatory and the Telescope Array detectors.
\end{itemize}

\section*{Acknowledgement}

This work was supported by the RSF Grant 22-12-00103.

\bigskip
\bigskip
\noindent {\bf DISCUSSION}

\bigskip
\noindent {\bf INGYIN ZAW:} Do these models reproduce the sky distribution of UHECR's at $E > 10^{20}$ eV?
If they come from the Galactic Center, they should point back to it, even if they are only a few.

\bigskip
\noindent {\bf ELENA ARBUZOVA:} 
Indeed, the arrival direction of the ultrahigh energy cosmic rays, if they originate from the Galactic Center, should indicate to the source. 
There are only one or maybe two events of such extremely high energy. One event may happen, even if  its probability is negligible.
According to the data the most energetic particle originated from the empty space in the sky without any visible source. 
This fact is in favour of our model of UHECR creation by heavy dark matter particle decays. 
On the other hand it may also come from ultra heavy particle annihilation in the Galaxy or from galactic dark matter halo.

\end{document}